\documentclass[sensors,article,submit,pdflatex,moreauthors]{Definitions/mdpi} 

\renewcommand{\linenumbers}{}

\firstpage{1} 
\pubvolume{1}
\issuenum{1}
\articlenumber{0}
\pubyear{2023}
\copyrightyear{2023}
\datereceived{ } 
\daterevised{ } 
\dateaccepted{ } 
\datepublished{ } 
\hreflink{https://doi.org/} 

\Title{SpecTracle: Wearable Facial Motion Tracking from Unobtrusive Peripheral Cameras}

\TitleCitation{SpecTracle: Wearable Facial Motion Tracking from Unobtrusive Peripheral Cameras}

\Author{Yinan Xuan $^{1}$, Varun Viswanath $^{1}$, Sunny Chu $^{1}$, Owen Bartolf $^{1}$, Jessica Echterhoff $^{1}$, and Edward Wang $^{1,}$*}

\AuthorNames{Yinan Xuan, Varun Viswanath, Sunny Chu, Owen Bartolf, Jessica Echterhoff, and Edward Wang}

\AuthorCitation{Xuan, Y.; Viswanath, V.; Chu, S.; Bartolf, O.; Echterhoff, J.; Wang, E.}

\address{%
$^{1}$ \quad University of California, San Diego\\
}

\corres{Correspondence: ejaywang@ucsd.edu}

\abstract{Facial motion tracking in head-mounted displays (HMD) has the potential to enable immersive "face-to-face" interaction in a virtual environment. However, current works on facial tracking are not suitable for unobtrusive augmented reality (AR) glasses or do not have the ability to track arbitrary facial movements. In this work, we demonstrate a novel system called SpecTracle that tracks a user's facial motions using two wide-angle cameras mounted right next to the visor of a Hololens. Avoiding the usage of cameras extended in front of the face, our system greatly improves the feasibility to integrate full-face tracking into a low-profile form factor. We also demonstrate that a neural network-based model processing the wide-angle cameras can run in real-time at 24 frames per second (fps) on a mobile GPU and track independent facial movement for different parts of the face with a user-independent model. Using a short personalized calibration, the system improves its tracking performance by 42.3\% compared to the user-independent model.}


\begin{document}

\begin{figure}[H]
  \includegraphics[width=\textwidth]{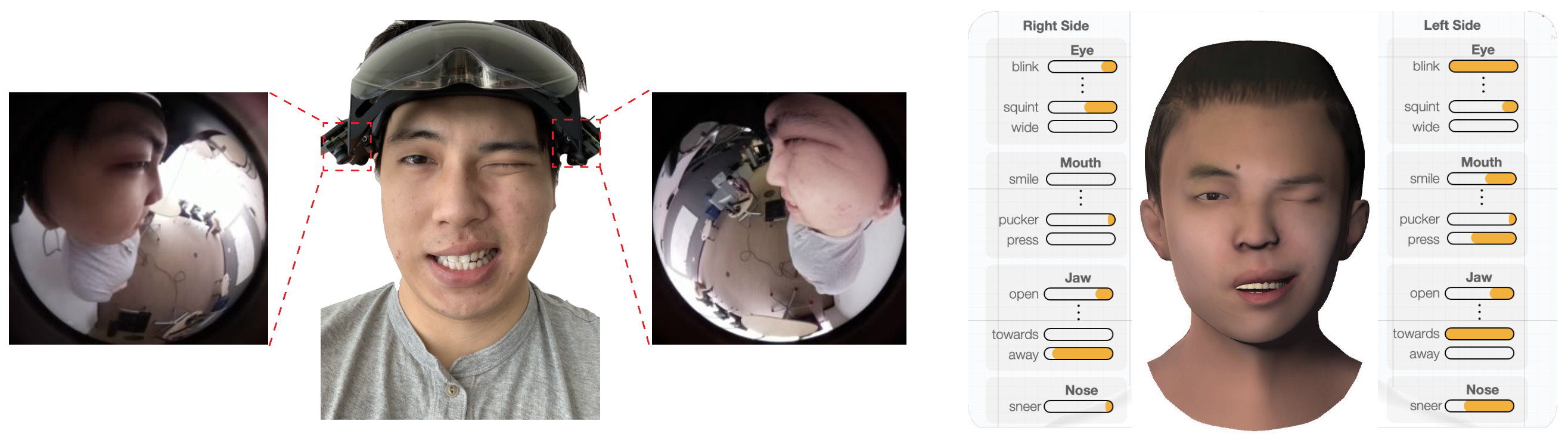}
  \caption{SpecTracle System}
  \label{fig:teaser}
\end{figure}
\section{Introduction}
Immersive face-to-face conversation is one of the most compelling use cases in Augmented Reality (AR). AR represents the next evolution in digital social interaction, allowing distant friends to appear just a few meters away in a shared 3D space. In order for AR to truly revolutionize the social space, elements crucial to communication must be faithfully recreated. AR systems must track a wearer's movements, postures, and most importantly facial motion to authentically replicate the user in a digital environment. To fully realize Augmented Reality's power to merge physical environments into digital interactive spaces seamlessly, AR systems must do this while striving to be fully portable. Instrumenting the environment with tracking hardware confines the user to a specific environment rather than augmenting any environment. As such, 'inside-out' tracking, or tracking just from a headset or other wearable device, is highly desirable, and it's essential to make the system  unobtrusive and low-profile to allow users to perform daily activities as naturally as possible.

Prior work in low-profile facial tracking systems has relied on the use of biosignals \cite{Gruebler2014EMGExpressions, chen2021exgsense, Verma2021ExpressEarSF, Wu2021BioFace3DC3} or acoustic interferometry \cite{Iravantchi2019Interferi} to classify the facial expression being made (e.g., smiling, frowning, etc.). However, these works only supported a limited number of expressions and cannot satisfy the need for immersive face-to-face communications. On the other hand, systems that can track a wide range of expressions require positioning cameras far in front of the face and pointing them directly at the wearer's face\cite{Li2015HeadMountedPerformance,Cao2016RealTimeAnimation, Laine2017ProductionLevelFacial}, which significantly hinders the users from a lot of actions.

In this work, we demonstrate SpecTracle, a novel facial tracking system that is compatible with AR visors and does not sacrifice wearability and mobility to deliver accurate facial tracking for AR face-to-face communication applications. 
SpecTracle relies only on two wide-angle lenses mounted flush to the periphery of the visor, making the setup viable for adaptation for AR systems. The system performs continuous motion tracking of facial movements for the eyebrows, eyes, nose, tongue, mouth, cheeks, and jaw. 

In addition, our method still benefits from the robustness of direct capture systems as it fully captures a frontal side profile of the face. Prior work that avoids having cameras in unwieldy locations has shown the use of cameras pointed forward from headphones to track the side contour of the face  \cite{Chen2020cface}. However, the system has yet to be demonstrated to function well in different environments and under free movement. Our evaluation demonstrates that SpecTracle can robustly track facial movement even when the user is walking or is otherwise involved in unconstrained free movement. 

In our user study, we recruited 9 participants with a mix of genders and facial skin tones. We demonstrate that the system is able to track the faces of all subjects, both indoors and outdoors. Moreover, we developed a personal calibration procedure that boosts system performance by 42.3\% in comparison to a user-independent model. This personal calibration requires the user to demonstrate a range of expressions, which takes only around 11 seconds. We demonstrate that the proposed algorithmic pipeline can run at 24.4 fps on a mobile GPU (iPhone 11) and 13.9 fps on previous generation mobile GPUs (iPhone 7).

\subsection{Contributions}
Our work addresses key issues that are critical to AR experiences. The main contributions of our work are summarized below: 
\begin{enumerate}
    \item We introduce an inside-out wearable camera facial motion tracking system for AR headsets while keeping the design unobtrusive.
    \item We present a facial tracking neural network model that can robustly track a new user's facial movement both indoors and outdoors during unconstrained movement.
    \item We offer a user-calibrated model that can improve facial tracking results by 42.3\% compared to the user-independent model. For the user-calibrated model's predictions at uncalibrated locations, the model performance improved by 28.3\% compared to the user-independent model.
\end{enumerate}

In the next section, we will examine the related work around the applications and methods of facial motion tracking. We will also discuss how our work improves the limitations presented in previous works.

\subsection{Applications of Facial Motion Tracking}
Facial tracking technologies typically fall under two different categories, recognition and motion tracking. Recognition is most often associated with the identification of an individual, as used with facial recognition, but can also be around the use of recognizing the emotion and affective state of an individual. In comparison, \textit{facial motion tracking}, which is the focus of this paper, is centered around tracking the movement of the face, such as the direction of eye gaze, blinking, mouth posture, head movement, and tongue movement. Face tracking has largely been a foundation for creating expressive facial movements and expressions for character animation in computer-generated graphics such as for movies and video games. \textit{Real-time character control} can also be made possible through facial motion tracking to enable virtual chats with avatars. This is a particularly interesting use case in the emergence of AR/ virtual reality (VR) where the presence of a person is embodied through a virtual face. Within the context of our work, we are particularly focused on the use of facial motion tracking in the domain of character control. 

\subsection{Facial Motion Tracking Technologies}
Motion capture (MoCap) has been widely used in computer-generated imagery productions to use as character animation input. Such control brings life to digital characters for pre-recording character movements for films and video games and real-time control of chat avatars. MoCap technologies can be divided into marker-based and markerless systems. Marker-based systems require placing physical patterns, usually in the shape of dots, on the actor's face and body. Invisible ink could also be used to hide the appearance, while tracking is done using IR light. This is in contrast with markerless systems, which rely directly on the sensing system to provide all the necessary tracking components. Such systems do not need the user to wear specialized clothing with markers. Instead, the user wears a device mounted typically on their head or stands in front of a camera system. Our proposed system, SpecTracle, is a markerless system for facial motion capture.

Marker-less face tracking, by and large, has been explored through the use of RGB color cameras facing the user. From a system setup standpoint, cameras are often placed about half an arm's length away. In a stationary situation, webcams have been used extensively to perform monocular tracking of the face for facial landmarks, blend shapes, and full mesh reproductions \cite{Li:2008:PUC:1358628.1358946, Thies2016Face2FaceRF}. In a mobile setup, cameras could be mounted on a headgear where the cameras are extended away from the body. Such a setup, which is aimed for use in actor facial expression capture in an open space, provides the ability for the actor to move around while the camera perspective on the face is fixed. Both monocular and multi-camera setups have been demonstrated to work well for facial tracking using an extended camera setup that can capture the full face. Although the dual-camera setup shown by \citet{Klaudiny2017MultiView} provides a similar concept to our side camera viewpoint, their setup is much further away from the face and placed at an oblique angle such that the front of the mouth is still captured. In our setup, we not only bring the camera far closer to the face (i.e. the lens is coincident with the glasses arm), the resulting camera angle is much more akin to a complete side profile of the face and requires inference of what the front of the mouth looks like. 

In contrast, IR camera-based solutions such as the Kinect \cite{Kinect} and PrimeSense facial recognition cameras on phones \cite{AppleARKit} largely employ the use of structured illumination on the face. As IR light is not conceived by the human eye or color cameras with IR filters, these patterned dot projections do not affect the visual qualities of the scene. The dot projection provides visual landmarks to calculate the depth of the surfaces in the scene, which gives a 2.5D depth map. Our current work does not exploit this class of sensing, however, a future direction can be to utilize a dot projection based system mounted in a similar fashion. 

\subsection{HMD Face Tracking Systems}
Most relevant to our work are head-mounted display (HMD) facial tracking systems that aim to replicate form factors of VR/AR/Mixed Reality (MR) headsets and ultra-low-profile smart glasses. The need to be truly wearable for everyday use is distinct from solutions being created for filming, which can be extended from the body. 

For VR HMDs, the challenge is to capture the facial expression while the face is occluded. \citet{Li2015HeadMountedPerformance} demonstrates a system that captures the upper face movement with strain gauges and the lower face with an RGB-D camera mounted on a selfie stick like an arm to look at the lower face. A series of works demonstrate a setup that uses multiple IR cameras: one at each eye and one at the front edge of the VR HMD pointing at the mouth \cite{Lombardi2018Deep, Wei2019Multiview}. This setup demonstrates successful tracking of the full facial motion tracking of the eyes, mouth, and jaw movements. 

Work has also been done to stay true to a low-profile form factor for wearable face tracking \cite{Scheirer1999ExpressionGlasses,Gruebler2014EMGExpressions}. Bernal et al. demonstrate a system called PhysioHMD that uses electrodes placed on an upper face mask similar in shape to a face mask that tracks a mix of electromyography and electro-dermal activity to classify between different facial expressions related to affective states \cite{ Bernal2018PhysioHMD}. \citet{chen2021exgsense} et al. also worked on a mask-shaped system using biopotential signals. In a similar face mask form factor, \citet{Iravantchi2019Interferi}'s uses acoustic interferometry to capture facial movements to track smiling gestures. Various works integrate IR reflection sensors around a glasses frame to capture surface contour movement as different parts of the face move closer and further away from different parts of the glasses frame \cite{Masai2016FaceExpressionPhotoReflective, Suzuki2017ReflectiveSensorinHMD, Cha2019FacialGestures}. Other work that aims to capture face-related movements instrumented the face through the ear \cite{Gajos2017CanalSense, Mark2017EarFieldSensing,Bedri2017Earbit}, and jaw movements from a necklace \cite{Zhang2020NeckSense}. 

While the works above are limited to classifying the user performing a set of full-face gestures rather than continuous tracking, more recent works have also explored low-profile systems that support continuous tracking. \citet{Elgharib2020Egocentric} proposes a system that can reconstruct a realistic frontal image of the user using a single camera dangling on the right side of the user's face. On the other hand, \citet{Chen2020cface} has developed C-Face, which can continuously track facial expressions via two cameras positioned at the ears and looking at the contour of the cheeks. A follow-up study also presents a neck-mounted wearable that allows continuous facial tracking \cite{chen2021neckface}. However, by virtue of the mounting location, neither system could be easily integrated into AR glasses.   

\subsection{Limitations of Current Works}
Although various approaches have been proposed to solve the facial tracking problem, in face-to-face chatting in AR context, a number of issues still remain unsolved. 

\begin{figure}[H]
  \centering
  \includegraphics[width=\linewidth]{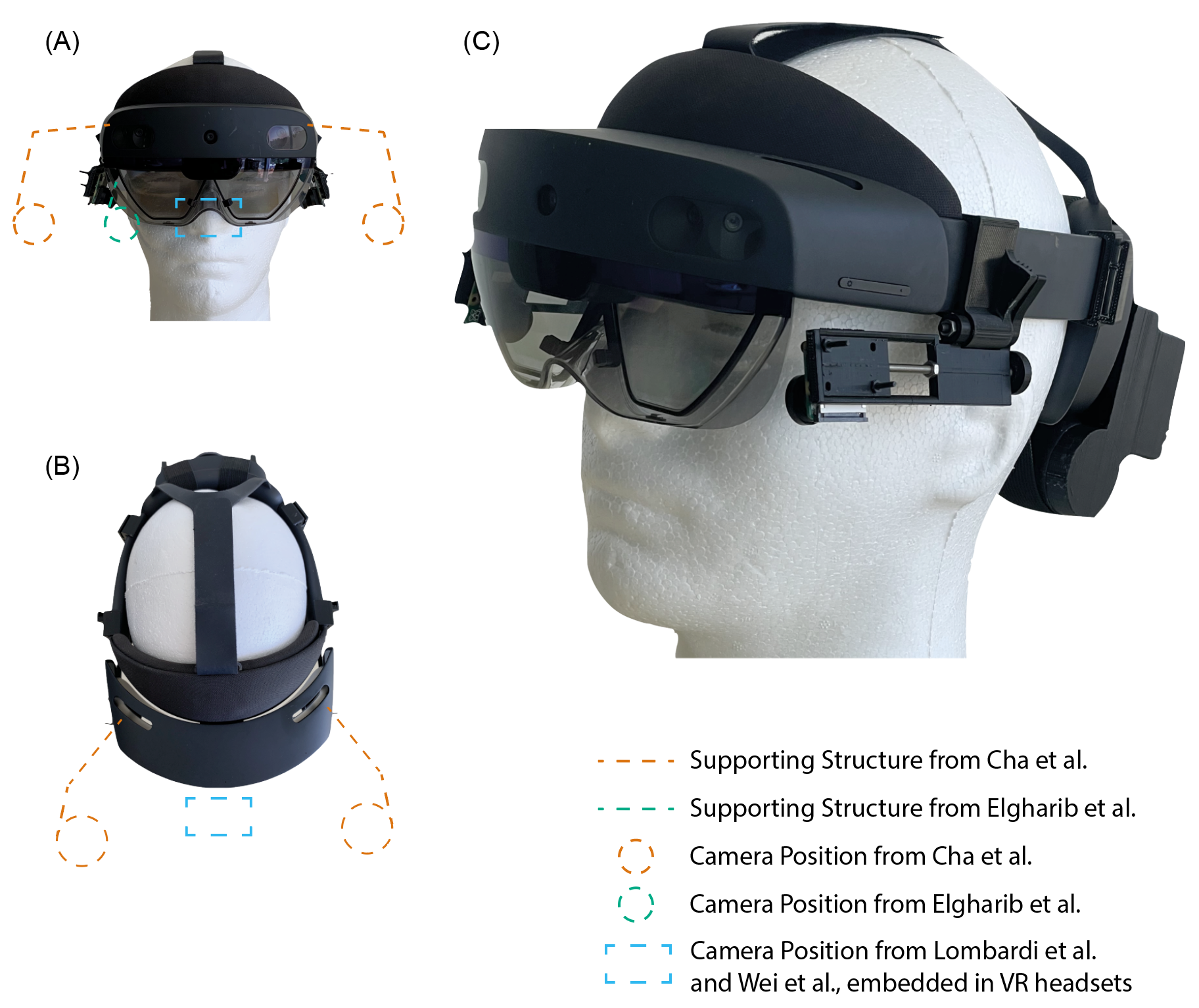}
  \caption{SpecTracle is unobtrusive and compatible with Hololens. Dashed lines and shapes indicate estimates of camera positioning from works of \citet{Cha2018towards}, \citet{Elgharib2020Egocentric}, \citet{Wei2019Multiview} and \citet{Lombardi2018Deep}. (A) Frontal view (B) Top view (C) Side view }
  \label{Hololens}
\end{figure}

The first issue is the portability of the HMD once mounted with cameras. In \citet{Lombardi2018Deep} and \citet{Wei2019Multiview}, the VR headset extends about 10cm from the face and the camera is positioned in front of the face, which is similar to a remote camera setup in nature. Similarly, in \citet{Cha2018towards}, the cameras are mounted on antenna-like structures which extend ~9 cm away from the face. This is not a feasible camera placement for low-profile setups such as AR glasses and smart glasses. To solve this issue, we design the system in an unobtrusive way, as shown in Figure \ref{Hololens}.

The second issue is the lack of user-independent solutions. While \citet{Lombardi2018Deep, Wei2019Multiview,Elgharib2020Egocentric} and \citet{Cha2018towards} can very well track facial expressions, the solution is highly calibrated to a single user. A new user would have to spend quite some time providing training data before using the system. In contrast, we provide a user-independent solution as an option for use cases where users do not have time to calibrate the system. In addition, we also provide a user-dependent solution should the user wish to spend a short amount of time on calibration to get a performance boost.

The third issue is the incompatibility to a complex background in mobile contexts. In  \citet{Chen2020cface}, although the solution is low-profile and user-independent, the system only works with users sitting in an unchanging background. We solve this problem by providing various training data so that our system is able to perform continuous full-face tracking while the user is walking around in both indoor and outdoor lighting conditions.

\section{Materials and Methods}

We present SpecTracle, an inside-out wearable facial motion tracking system using two ultra-wide-angle RGB cameras to track the continuous facial motions of the wearer. We envision our system to be used for digital face-to-face conversations in the AR context, much like the way video chats work today for 2D displays. Our unobtrusive hardware design and software solution allow the user to engage in conversations in the form of virtual characters while being unconstrained to move around indoor and outdoor environments. 

\subsection{Hardware}

\begin{figure}[H]
  \centering
  \includegraphics[width=\linewidth]{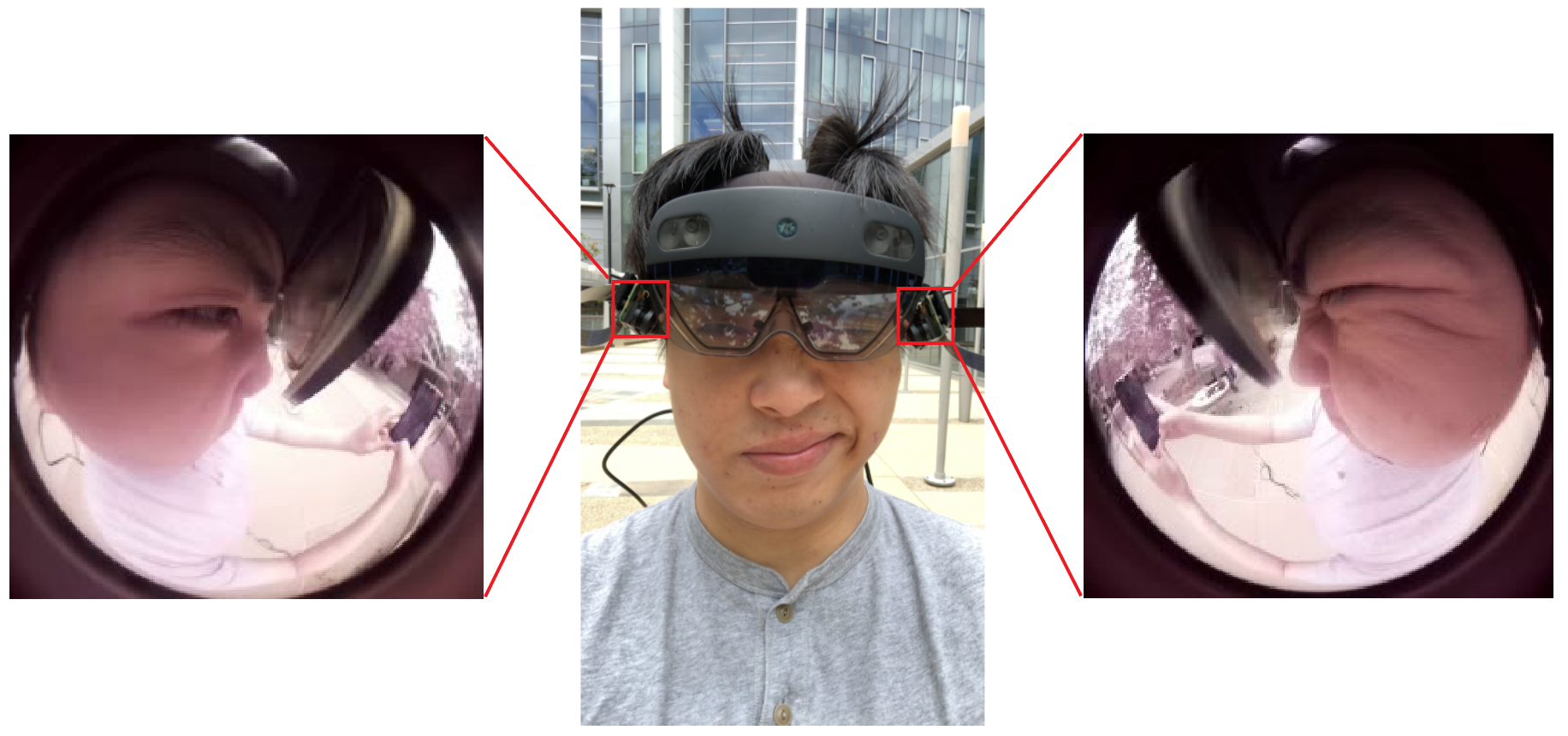}
  \caption{Example of Images captured by SpecTracle's fisheye lens cameras.}
  \label{camera_view}
\end{figure}

The SpecTracle hardware consists of two cameras captured synchronously by a StereoPi (Figure \ref{Hololens}). 3D printed clip mounts fix the cameras and the StereoPi on a Hololens. The cameras are positioned at the ideal angle (40$^{\circ}$) and position (right next to the visor), determined through a small user study. In order to position the camera flush with the visor, which is about 1 inch from the face (Figure \ref{camera_view}), we use a 220$^{\circ}$ fisheye lens \footnote{https://www.entapano.com/shop-en/product/raspberry-pi-lens-220/}. 

\subsection{Machine Learning}
\begin{figure}[H]
  \centering
  \includegraphics[width=\linewidth]{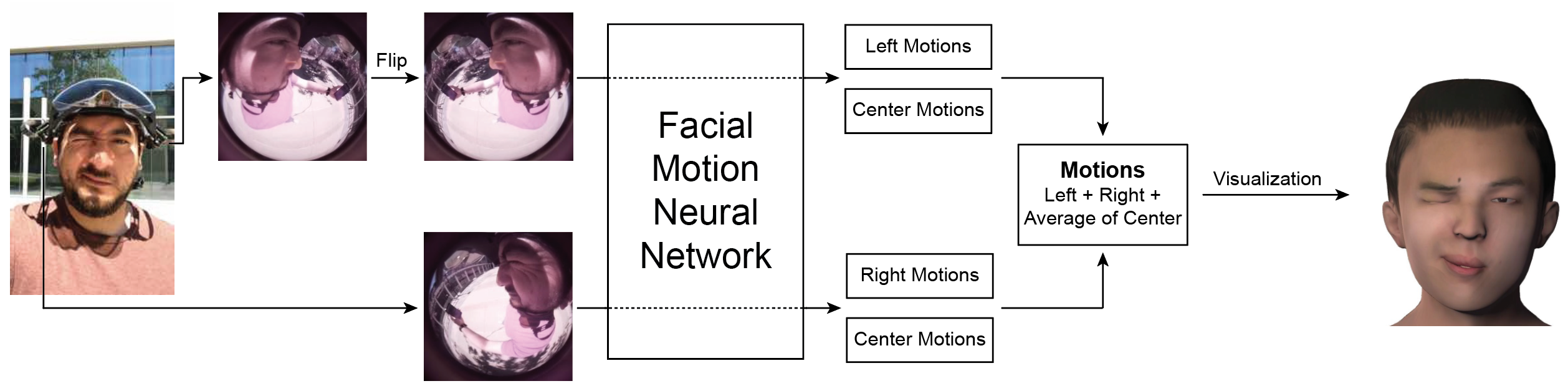}
  \caption{Data flow diagram of the facial tracking system. The left and right camera feed into the neural network to predict the corresponding left and right blend shape values. The center blend shapes are averaged.}
  \label{data_flow}
\end{figure}

Our machine learning model takes the side-view images from the wide-angle cameras as input and predicts a set of blend shape values that represent facial motion. (Figure \ref{data_flow}) After describing our blend shapes, we will discuss our neural network architecture.

\subsubsection{Facial Representation Using Blend Shapes}

A blend shape describes a particular deformation of a 3D object and is frequently used to control virtual animated objects such as a face. The scalar value associated with each blend shape describes the expressiveness of the deformation, and multiple blend shapes can be animated simultaneously, allowing for continuous and additive expressions of various 3D motions. By building our system to support a much higher degree of freedom, our system is not limited to set combinations of facial gestures. This is inherent to camera-based setups where the eyebrow, eyes, mouth, jaw, cheeks, tongue, and nose are spatially decoupled and thus can all be tracked independently. The use of blend shapes is a standard for character animation, is an output format for many face tracking systems \cite{AppleARKit, Pinscreen} and has also been used as the representation in other novel tracking systems in research \cite{Klaudiny2017MultiView}. SpecTracle uses a set of 52 blend shapes that are compatible with the Apple iOS ARKit Face Tracking software \cite{AppleARKitBlendshape}.

The utilization of blend shape in this work is crucial for the system to provide full facial articulation that has been worked on in the graphics domain. A critical part of enabling our system to track blend shapes over pre-determined facial expressions is the way we captured the training data. Traditionally, training data for open-frame systems has been based on asking subjects to perform a set of expressions many times to provide the machine learning classifier training data. Instead, we co-register the subject with a frontal-facing face tracking system (iPhone ARKit) to provide frame-by-frame blend shape references for system training. This provides our neural network with training data of the blend shape values of the current facial expression for each image sample. 

This frame-by-frame capture approach not only provides more granular articulation of different facial features independently for the two sides of the face, but also provides far more data. In our user study of 9 subjects, we collected a total of 180 minutes of facial movement data. Each frame acts as an independent data point to the neural network system, providing us with a total of about 172k samples.

Our blend shapes control different facial features (eyebrow, eye, nose, cheek, mouth, jaw, and tongue) for both sides of the face. Of the 52 total blend shapes we use, 18 blend shapes control the left side of the face, 18 other blend shapes control the right side of the face, and 16 blend shapes control the center of the face. Our side-specific blend shapes control facial features that skeletally can be moved independent of the other side of the face (eye blinking, corner of the mouth movements), while our center blend shapes (such as jaw motion mouth-puckering) are shared. Specifically, our system uses 14 eye and eyelid blend shapes(7 L/R), 4 jaw blend shapes (4 C), 23 mouth blend shapes (7 L/R + 9 C), 5 eyebrow blend shapes (2 L/R + 1 C), 3 cheek blend shapes (1 L/R + 1 C), 2 nose blend shapes (1 L/R), and 1 tongue blend shape. Table \ref{tab:blendshapes} lists all the blend shapes used in our system. Each of our blend shapes is a continuous value between 0 and 1, with 0 being a neutral state and 1 being a fully activated state. For example, when all values are set to 0, the face would have a neutral expression, with no smile, eyes open and looking forward, and no cheek or jaw motion. If \textit{jawOpen} is at 1, the jaw is fully open. If, in addition, the \textit{eyeBlinkLeft} is at 0.5 and the \textit{eyeBlinkRight} is at 0, then the eye of the user will be half closed on the left and fully open on the right, while the jaw remains open. Since blend shapes are additive, the multitude of dimensions of the mouth shape helps to create a wide variety of expressive mouth shapes beyond simple open and close. For example, to describe a mouth that is stretched to the right and slightly open, both the \textit{mouthStretchRight} and the \textit{jawOpen} blend shapes can be activated simultaneously (Figure \ref{fig:teaser}). 

\begin{table}
  
  \begin{tabular}{ m{1.5cm} | m{11cm} } 
  \toprule
 \textbf{Part} & \textbf{blend shapes}   \\
 \midrule
\textbf{Eye} & eyeBlink*, eyeLookDown*, eyeLookIn*, eyeLookOut*, eyeLookUp*, eyeSquint*, eyeWide*  \\
\midrule
\textbf{Mouth and Jaw} & jawForward, jawLeft, jawRight, jawOpen, mouthClose, mouthFunnel, mouthPucker, mouthLeft, mouthRight, mouthSmile*, mouthFrown*, mouthDimple*, mouthStretch*, mouthRollLower, mouthRollUpper, mouthShrugLower, mouthShrugUpper, mouthPress*, mouthLowerDown*, mouthUpperUp*, tongueOut   \\
\midrule
\textbf{Eyebrows, Cheeks, and Nose} & browDown*, browInnerUp, browOuterUp*, cheekPuff, cheekSquint*, noseSneer*   \\

             \bottomrule
\end{tabular}
\\ \hskip 6.5cm \small * has Left/Right for each side of the face. (e.g. eyeBlinkL, eyeBlinkR)\\
\caption{List of the specific blend shapes we implement in SpecTracle. The facial motions can be inferred from the blend shape names.}
\label{tab:blendshapes}
\end{table}

\subsubsection{Facial Motion Neural Network}
SpecTracle uses a neural network model to predict its facial blend shapes from a single image frame from one side of the face. The network outputs a total of 34 values, with 18 values describing the side blend shapes corresponding to the side of the face where the image is from and 16 values describing the center blend shapes. To get the 52 blend shapes describing the entire face, we (1) make predictions on the right and left images respectively, (2) concatenate the left and right side blend shape predictions, (3) average the two predictions for the center blend shapes, and (4) return the left, right, and averaged center blend shapes as the final predictions of the system. (Figure \ref{data_flow})

The model is trained using transfer learning, starting from a VGG19 model \cite{simonyan2014very} pre-trained on ImageNet with pre-trained weights downloaded from the TensorFlow library \cite{tensorflow2015-whitepaper}. The out-of-the-box pre-trained VGG model is modified so that its last dense layer is replaced with one single dense layer that outputs 34 values (18 side blend shapes and 16 center blend shapes). To constrain the predicted value within the range of 0 to 1, a sigmoid activation function was added after the last layer of the network. Contrary to many other transfer learning works, we do not freeze any layers and train all the weights whenever we train the model. We experimented with freezing the weights but found that doing so weakened performance. Other than VGG, we have also explored ResNet \cite{he2016deep} and AlexNet \cite{krizhevsky2012imagenet} network architectures. Among all three network architectures, we chose VGG because VGG yielded the best tracking result. (The VGG outperforms the ResNet by 1.8\% and AlexNet by 9.8\%)

We chose to define the prediction task such that the model only uses an image of one side of the subject's face with the goal of increasing dataset variance and simplifying the model input space. In order to reduce the complexity of the training data and increase the number of samples in our dataset, the left side of the face is horizontally flipped so that we only train on faces facing one direction. This way, the model only needs to learn features for one single face orientation. In addition, by using both images as inputs, the neural network may adapt better to variance introduced by the camera positioning, as the HoloLens is not always perfectly symmetrically placed on the user's head. The images are then resized and normalized before being fed into the neural network model.

\subsubsection{Data Enhancement}
Our data collection happened both indoors and outdoors, with various lighting conditions. Thus, the camera automatically adopted various white balancing settings throughout the whole data collection process. To make our model robust to various white balance settings, We used WB color augmenter \cite{Afifi_2019_ICCV} to randomly change the white balance of our data. 

\subsubsection{Loss Function}
Due to how our blend shapes describe facial expressions, only a few blend shapes are activated at any one time. As such, most of the time, most blend shape values will be at 0 (neutral) and only one or a few blend shapes are activated (has a high value) when the face deviates from neutral. 
Therefore, when treated as a feature, most blend shape data are heavily imbalanced, with most values close to or at 0. 
When we used mean absolute error with no modifications as the loss function, our model predicts everything to be 0. To account for the imbalanced distribution, the system employs a custom loss function where the loss is scaled for each blend shape in each sample. Specifically, \(loss = |prediction - label| * 50^{label}\), and the scale factor is determined empirically. As a result, the model would be heavily penalized when it should have predicted a high value but instead predicted a low value. This resulted in a far more responsive prediction of non-neutral states. 

\subsection{Avatar Visualization}
Visualizing our system's blend shapes required a virtual avatar with facial rigging consistent with our blend shape dimensions. Because our blend shapes are consistent with the Apple iOS ARKit blend shapes, our system can control any characters rigged with the ARKit blend shapes, such as the Animoji characters.

\subsection{Data Collection}
We performed a series of studies to gather data that we use to make the best design choices, and evaluated the performance of our system. We will describe how we collected the data and how we used it to improve and evaluate our system.

\subsubsection{Protocol}
During each data collection session, the subject was asked to perform a series of predefined facial expressions in a randomized order for 10 rounds. 1 clip of video was recorded each round. Each round is 2 minutes and splits into three sub-rounds of 40 seconds each, where the sub-rounds had the subject do (1) eye and eyebrow motions, (2) mouth and jaw motions, and (3) cheek, nose, and tongue motions. Subjects were asked to walk around slowly in a \textasciitilde500 square feet area. Each subject repeats 5 rounds of data collection in one indoor and one outdoor environment, constituting a total of 10 rounds per subject. The procedure has been approved by University of California, San Diego's IRB under Protocol \#200201XL. 

\begin{figure}[H]
  \centering
  \includegraphics[width=\linewidth]{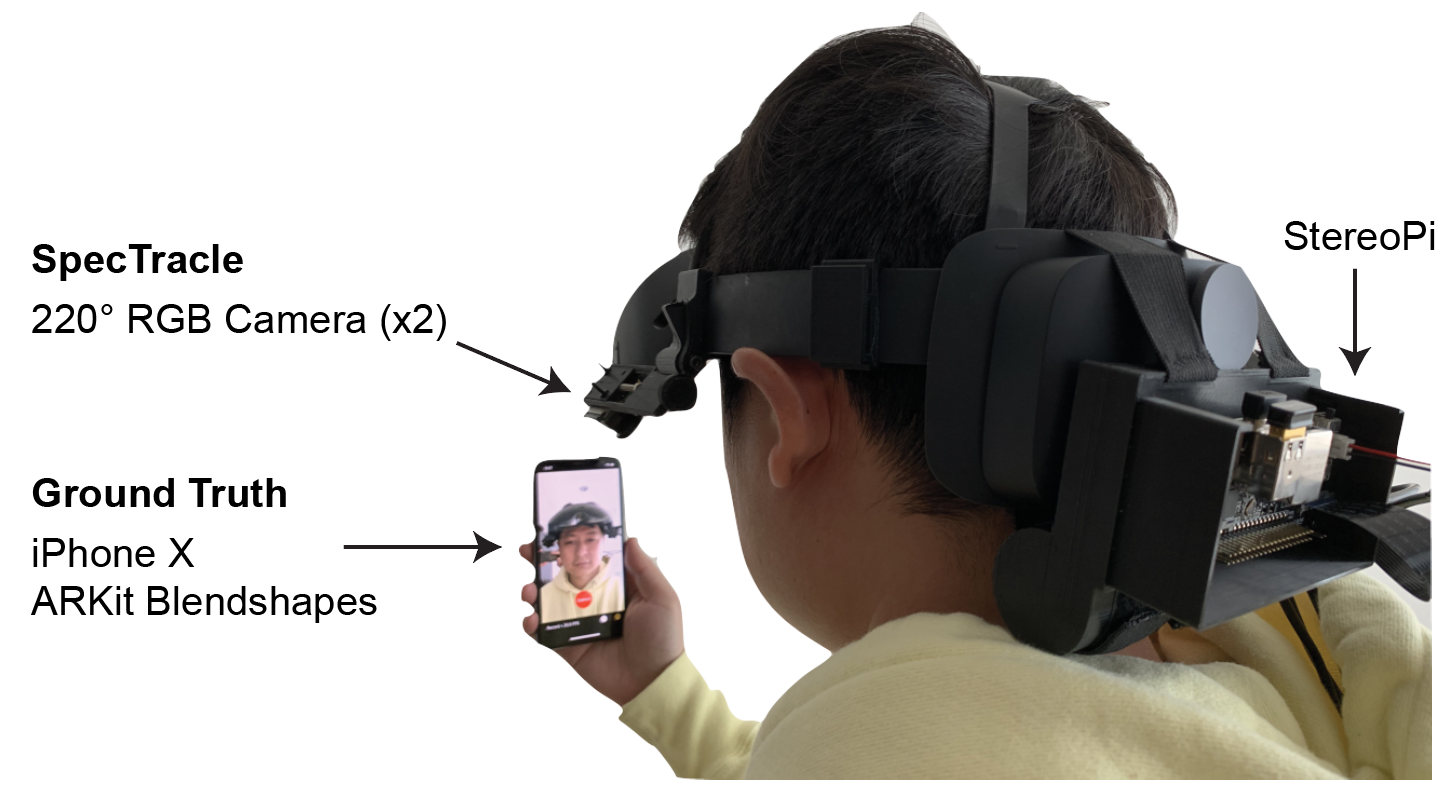}
  \caption{Data collection setup.}
  \label{data_collection}
\end{figure}

\subsubsection{Setup}
\textbf{StereoPi Data} To collect the data, each subject was asked to wear the setup. After putting on the SpecTracle devices, the subjects were asked to adjust the headset positioning to make sure the images captured by both cameras were as symmetrical as possible. The visor of the Hololens was flipped up to ensure proper ground truth collection. Subjects with long hair that covered the forehead were asked to show the eyebrows clearly to ensure the ground truth tracking works properly. Overall, we deemed the process to be as natural as possible. The cameras recorded videos at \textasciitilde8 frames per second.

\textbf{Ground Truth} We built a custom application on an iPhone X that uses the iPhone's IR face tracking system to co-register the values of the 52 blend shape values frame-by-frame. The custom application was modified from an open-source application that accesses blend shape data using the ARKit SDK \footnote{https://github.com/elishahung/FaceCaptureX}.

During each data collection session, subjects were asked to hold the iPhone in front of their faces to record a video of their facial motions. Subjects were asked to keep the entirety of the front of their face within the field of view of the front-facing iPhone camera, to ensure their faces were properly detected. The video was recorded at 30 frames per second and, in each frame, the set of Apple blend shape values was extracted from the subjects’ facial expressions. Because the iPhone would return blend shape values based on whatever expression the subject is making, subjects were not required to perform the exact predefined facial expression. Furthermore, all frames between target expressions were also recorded and had intermediate values. As the protocol randomized the order and subjects were not asked to return to a neutral face between expressions, we were able to increase the diversity of our data. This method of acquiring ground truth enables our use of neural networks as it provides us with a large number of labeled instances (i.e. each frame is a sample).

\subsection{Dataset}

\subsubsection{Data Preparation}

\textbf{Synchronizing Data from Two Sources}\quad Since facial expressions, such as blinking, can take place in 100 milliseconds \cite{Kwon2013High-speed}, it is crucial to have data synchronization. Although the two cameras on the Raspberry Pi were synchronized, the recordings between the iPhone and Raspberry Pi were not synchronized. Therefore, timestamps of all frames were saved to help synchronize the data in recordings from both the iPhone and Raspberry Pi. However, because the system time on Raspberry Pi had a slight offset from that on the iPhone, we needed to manually adjust the offset using eye blinking as guidance.

\textbf{Data Cleaning}\quad We remove the following data from our dataset:
  \begin{itemize}
      \item Data where the side video frame cannot be synchronized to the ground truth frames due to their difference in frame rates.
      \item Data where the side video frames are the first 3 frames of each recording, which often has large fluctuations of exposure settings.
      \item Data where the ground truth is not properly recorded due to subjects accidentally moving their faces out of the field of view of the iPhone's front camera, causing the iPhone's failure to detect their faces.
  \end{itemize}

\subsubsection{Subjects}
We recruited 9 subjects (3F/6M) who participated in a data collection study, with 6 subjects (2F/4M) in the Fitzpatrick Skin Type \cite{sachdeva2009fitzpatrick} range of I-III (lighter tones) and 3 subjects (1F/2M)in the Fitzpatrick Skin Type range of IV-VI (darker tones). 
With a total of 9 subjects, after data cleansing and preparation, our dataset resulted in a total of 172k samples.

\section{Results}

For each participant, we trained two neural network models: the user-independent model and the user-calibrated model. The \textit{user-independent model} is able to track a new user's face, while the \textit{user-calibrated model} can boost the tracking accuracy even higher when tracking a calibrated user's face. 
\\
\\
In this section, we are going to first introduce the way we evaluated our models' performance. Then discuss each model's training/testing scheme and testing results.

\begin{figure}[H]
  \centering
  \includegraphics[width=\linewidth]{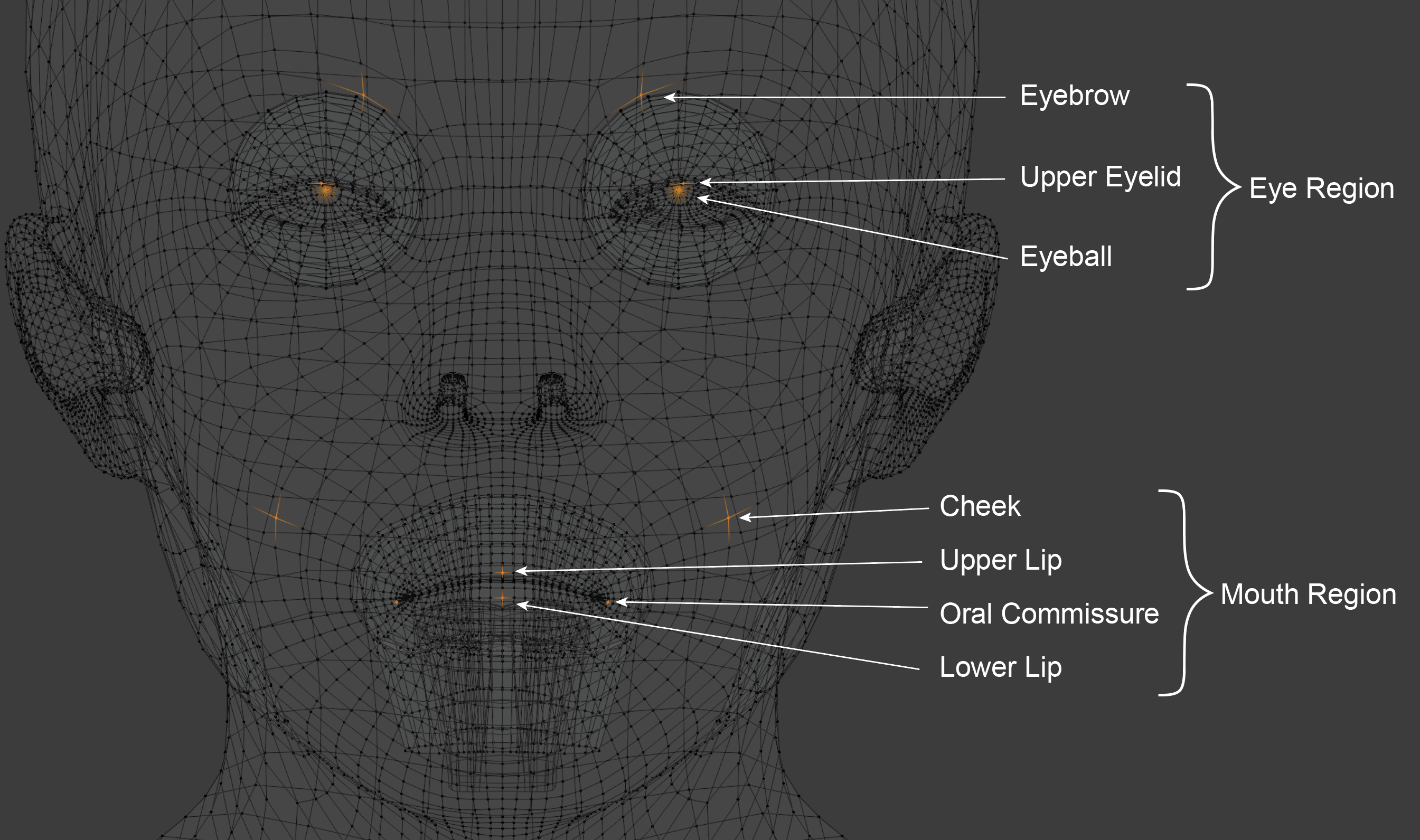}
  \caption{Vertices chosen for error evaluation. A total of 13 vertices are chosen (a vertex at the tip of the tongue is not shown).}
  \label{blender_points}
\end{figure}

\subsection{Model Performance Evaluation}
To examine a model's performance, we tracked the 3D coordinates of specific vertices from a realistic 3D face model mesh (Figure \ref{blender_points})\cite{blender}. This evaluation is favored over comparing the predicted blend shape values directly to the labels because the blend shape values are in relative scales and the error does not reflect the actual projection onto facial motion. Also, as previously mentioned in section 2.2.4, the data is heavily imbalanced, with most values being zeros. Therefore, a model making every prediction to be around zero would also have a small error if the model is evaluated by simply comparing the predicted values and the labels. To perform our evaluation, a face was animated with the ground truth label's blend shape values and also with the prediction's blend shape values. The error was then measured as the 3D Euclidean distance between a vertex from the ground truth label animated face to the one by the prediction. The unit of distance was converted to millimeters with the average inner canthal distance \cite{abdullah2002inner} as a reference.

\begin{figure}[H]
  \centering
  \includegraphics[width=\linewidth]{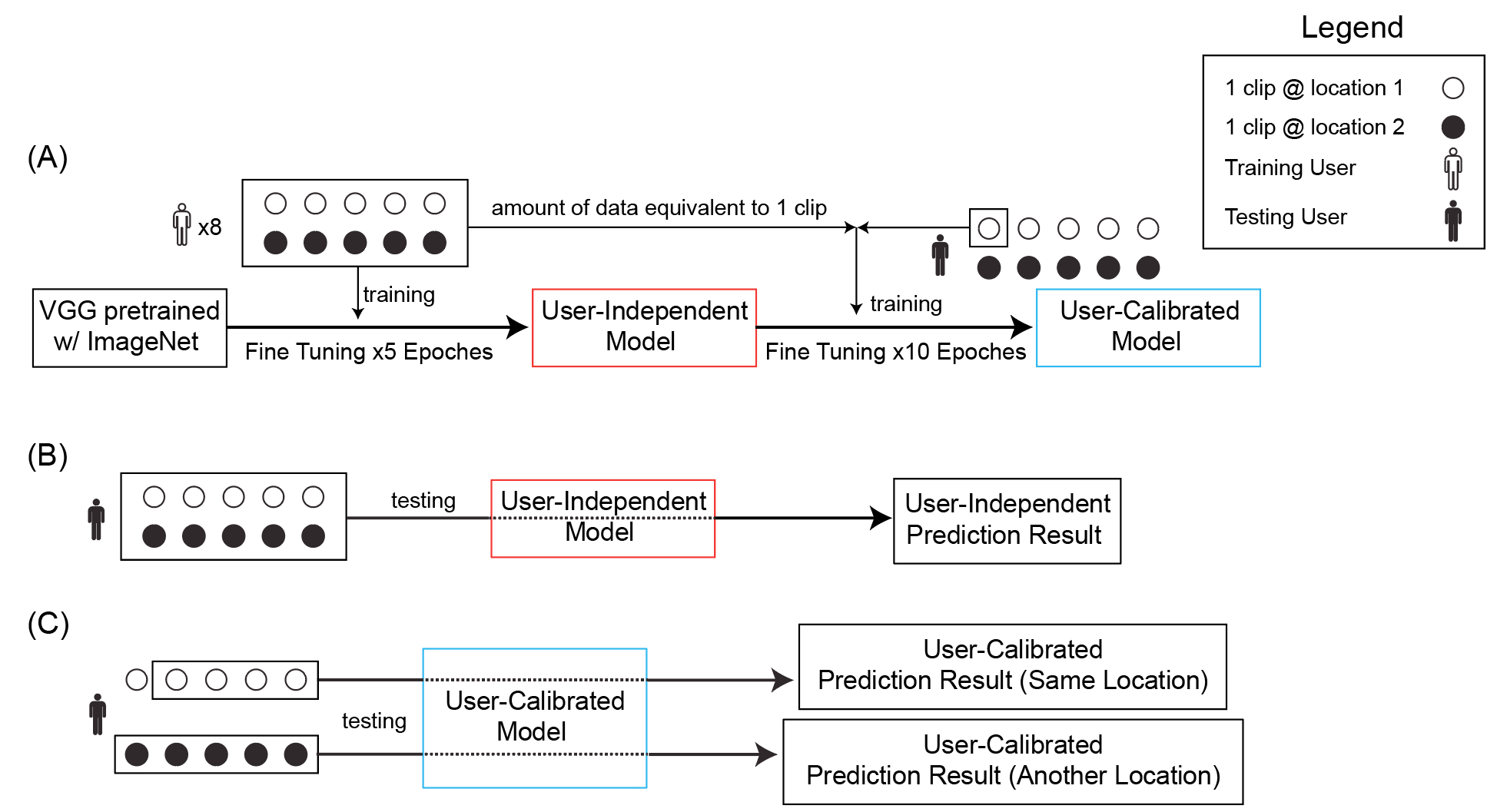}
  \caption{Training and Testing Scheme. Each participant is selected to be the testing user in turn, while the remaining 8 participants are the training user. Each user has 5 clips of data recorded at location 1 (indoor) and 5 clips of data recorded at location 2 (outdoor). (A) Training Scheme: all training user's data is used to fine-tune the VGG model for 5 epochs, generating a user-independent model. One clip of testing the user's data is then randomly selected and combined with the same amount of data randomly sampled from all training user's data to fine-tune the user-independent model for 10 epochs, generating a user-calibrated model. (B) All testing user's data is used to test the user-independent model. (C) The remaining data from the testing user, which is recorded at the same location as the data used for training the user-calibrated model, is used to test the user-calibrated model's performance in a calibrated location. The remaining data from the testing user, which is recorded at another location, is used to test the user-calibrated model's performance in an uncalibrated location.}
  \label{scheme}
\end{figure}

\begin{figure}[H]
  \centering
  \includegraphics[width=\linewidth]{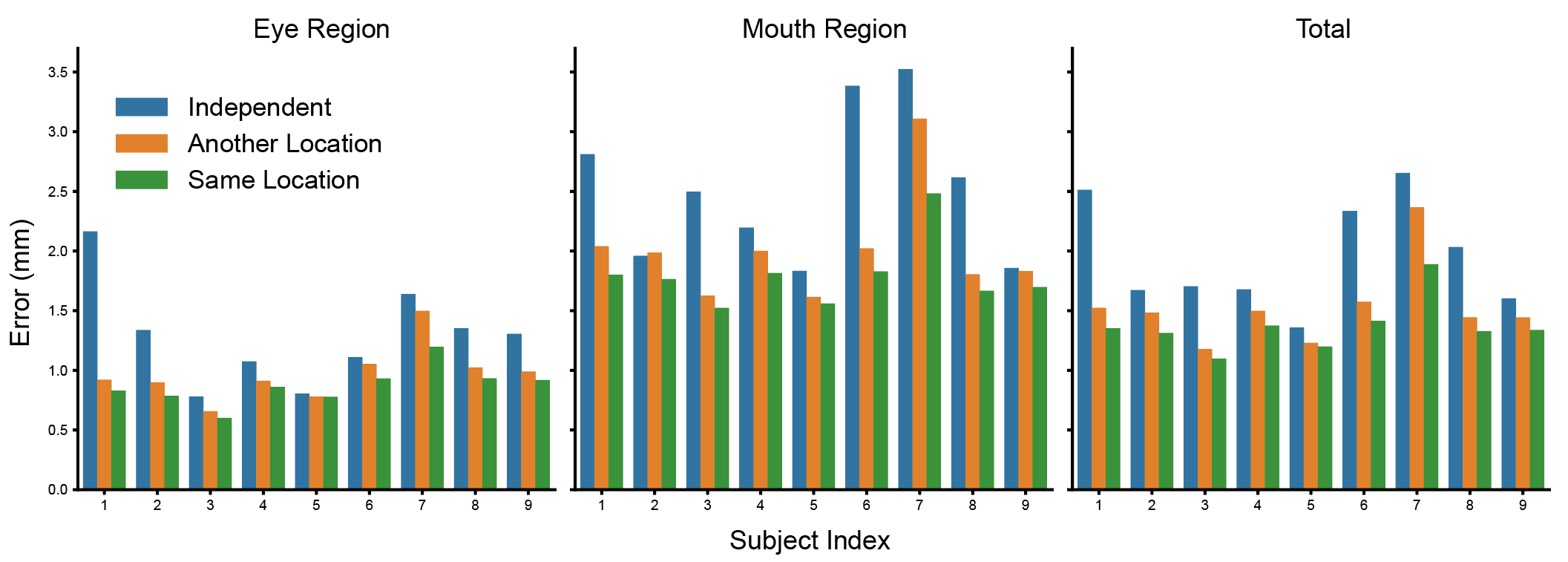}
  \caption{Model's Performance. Legend: "Independent" is the performance of the  user-independent model; "Another Location" is the performance of the user-calibrated model that is fine-tuned by the user's data collected in a different location from the data used for testing; "Same Location" is the performance of the user-calibrated model that is fine-tuned by the user's data collected in the same location as the data used for testing.}
  \label{error_plot}
\end{figure}

\subsection{User-Independent Facial Motion Tracking}

We started by examining how a user-independent model performs on a subject whose data had never been introduced to the model during training. This mimicked how a factory-calibrated model where the system was already trained on a variety of data but had never seen the user's own face.

To train the user-independent model, we first selected 1 subject as the testing user. The remaining 8 subjects were designated to be the training users. The user-independent model was a result of 5 epochs of fine-tuning from a VGG19 model pre-trained with ImageNet (Figure \ref{scheme}.A).

To examine the model's performance, we tested the \textbf{independent} model with data from the testing user.(Figure \ref{scheme}.B) The user-independent model achieved an average error of 1.29$\pm$0.14 (mm) for the 6 vertices tracked at the eye region and 2.52$\pm$0.21 (mm) for the 7 vertices tracked at the mouth region. Overall, the error was 1.95$\pm$0.15 (mm) for all tracked vertices. (Figure \ref{error_plot})

\subsection{User-Calibrated Facial Motion Tracking}

Next, based on the user-independent model, we further fine-tuned a user-calibrated model for each testing user. One can imagine the data could be quickly captured by having the user calibrate with a short sequence of facial movements. This mimics a user calibration process when a user first sets up the device for use, much like adding a new face identity, voice, or fingerprint on smartphones. To train the user-calibrated model, the user-independent model was fine-tuned for an additional 10 epochs. 50\% of the data used to fine-tune the model was randomly sampled from one randomly chosen clip of the testing user's data, while the other 50\% was randomly sampled from the data used for the user-independent model training (Figure \ref{scheme}.A).

\begin{figure}[H]
  \centering
  \includegraphics[width=8.5cm]{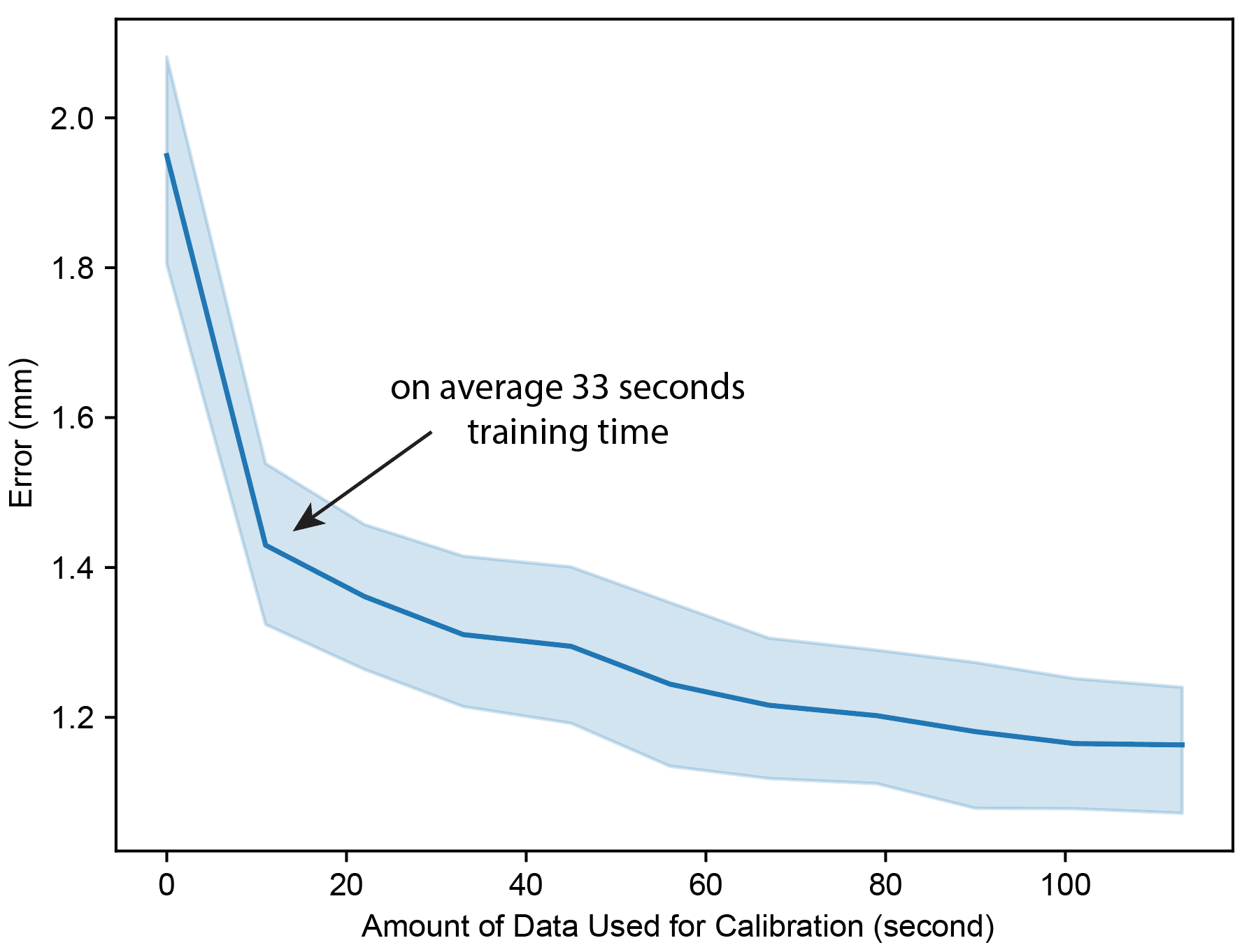}
  \caption{Calibration Curve. The error decreases as more data from the testing user are used for calibration. The shaded region indicates the standard error of the mean. The unit of the amount of data is converted to seconds based on 8 frames per second camera frame rate.}
  \label{calibration_curve}
\end{figure}

We first investigated the relationship between the amount of data used for calibration and the error to inform the amount of personalization data needed. We did so by randomly sampling various fractions of the calibration clip from one of the testing user's clips (each clip is approximately 2 minutes). The error decreased as more data is used for calibration (Figure \ref{calibration_curve}), but we saw that much of the improvement was met at about 10\% of the data (approximately 11 seconds). Across subjects, on average, the error was reduced by 26.4\% when calibrated with 11 seconds of calibration data. The model tuning took about 30 seconds on an RTX 2080Ti GPU. Therefore, we used this amount of data to generate all our following results. 

To make the evaluation realistic, we assumed that the user would only calibrate the system once, even when they were going to use the system later in a new environment. In our case, we collected data in one indoor and one outdoor location. Thus, we evaluated the system in two settings: (1) the testing data are the remaining data from the same location as the calibration data (\textbf{same location}), and (2)  the testing data are the ones collected in a location different from the calibration data (\textbf{another location}). (Figure \ref{scheme}.C) 

For the user-calibrated model tested in \textbf{another location}, the average error was 0.97$\pm$0.08 (mm) for the 6 vertices tracked at the eye region and 2.00$\pm$0.15 (mm) for the 7 vertices tracked at the mouth region. Overall, the error was 1.52$\pm$0.11 (mm) for all tracked vertices, which increased by 28.3\% compared to the error of the user-independent model.(Figure \ref{error_plot})

For the user-calibrated model tested in the location seen during calibration (\textbf{same location}), the average error was 0.87$\pm$0.05 (mm) for the 6 vertices tracked at the eye region and 1.79$\pm$0.09 (mm) for the 7 vertices tracked at the mouth region. Overall, the error was 1.37$\pm$0.07 (mm) for all tracked vertices, which increased by 42.3\% compared to the error of the user-independent model.(Figure \ref{error_plot}) 

After the t-test, we found that the calibration, even tested in \textbf{another location}, had significantly improved the model's performance (p-value = 0.02). While the error was lower when tested in the \textbf{same location} compared to \textbf{another location}, the errors were not significantly different (p-value = 0.25). 

\begin{figure}[H]
  \centering
  \includegraphics[width=\linewidth]{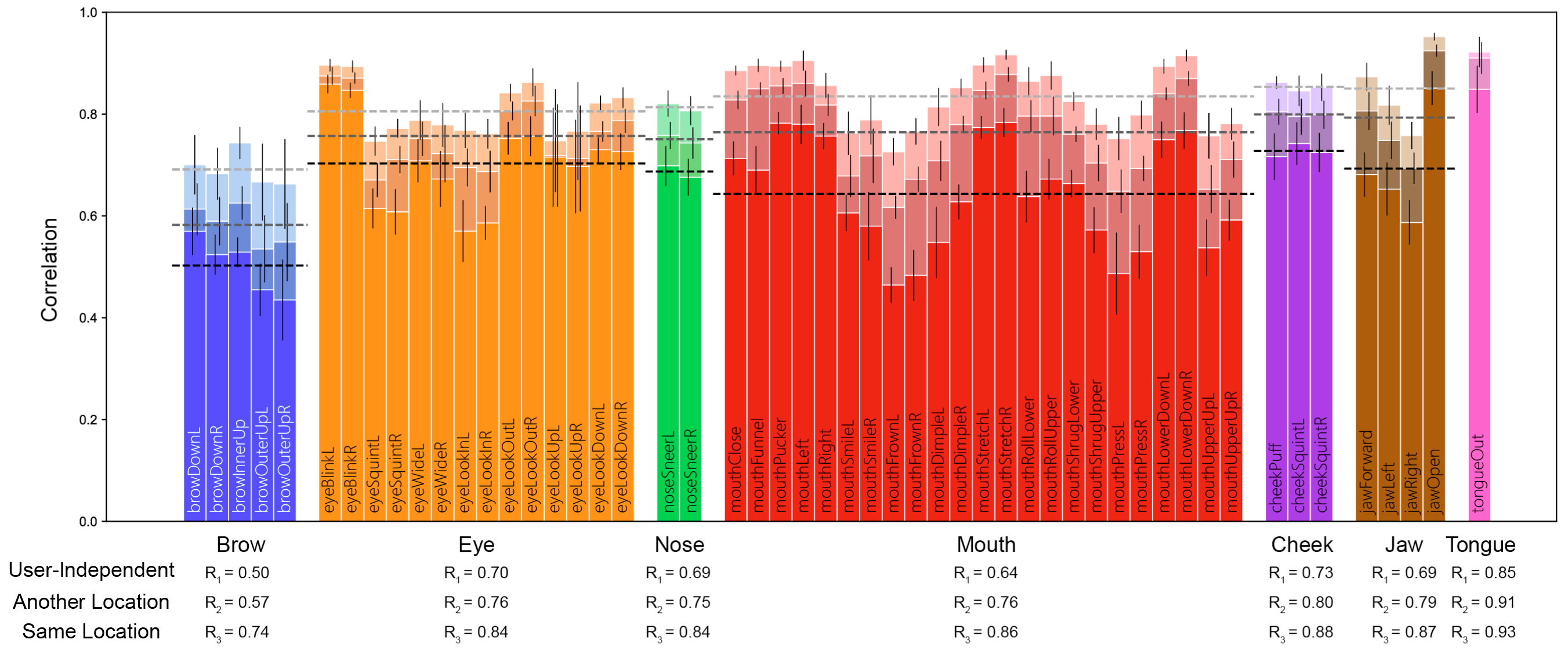}
  \caption{Comparison of models' performance across all 52 Blendshapes. The lower dashed line is the average correlation value across subjects of the user-independent model. The middle dashed line is for the user-calibrated model tested in another location. The upper dashed line is for the user-calibrated model tested in calibrated location. Below the barplot is a summary of each model's performance for different facial regions, averaged across blend shapes related to the corresponding region.}
  \label{52BS}
\end{figure}

Other than the reconstruction error, we also calculated the Pearson correlation of each blend shape between the predicted value and the reference value from the iPhone  (Figure \ref{52BS}). For the user-independent model, the correlation coefficient on average is R=0.66 and the correlation for the different body parts are as follows: Brow (R=0.50$\pm$0.05), eyes (R=0.70$\pm$0.04), nose (R=0.69$\pm$0.04), mouth (R=0.64$\pm$0.04), cheeks (R=0.73$\pm$0.04), jaw (R=0.69$\pm$0.04), and tongue (R=0.85$\pm$0.05). 

For the user-calibrated model when tested in another location, the correlation coefficient averaged to be 0.75 with R values for the different body parts as follows: Brow (R=0.57$\pm$0.05), eyes (R=0.76$\pm$0.04), nose (R=0.75$\pm$0.03), mouth (R=0.76$\pm$0.03), cheeks (R=0.80$\pm$0.03), jaw (R=0.79$\pm$0.03), and tongue (R=0.91$\pm$0.03), shown as deltas on Figure \ref{52BS}. 

For the user-calibrated model when tested in the calibrated location, the correlation coefficient averaged to be 0.82 with R values for the different body parts as follows: Brow (R=0.74$\pm$0.06), eyes (R=0.84$\pm$0.03), nose (R=0.84$\pm$0.03), mouth (R=0.86$\pm$0.02), cheeks (R=0.88$\pm$0.02), jaw (R=0.87$\pm$0.02), and tongue (R=0.93$\pm$0.03), shown as deltas on Figure \ref{52BS}. 

\subsection{Qualitative Results}

\begin{figure}[H]
  \centering
  \includegraphics[width=\linewidth]{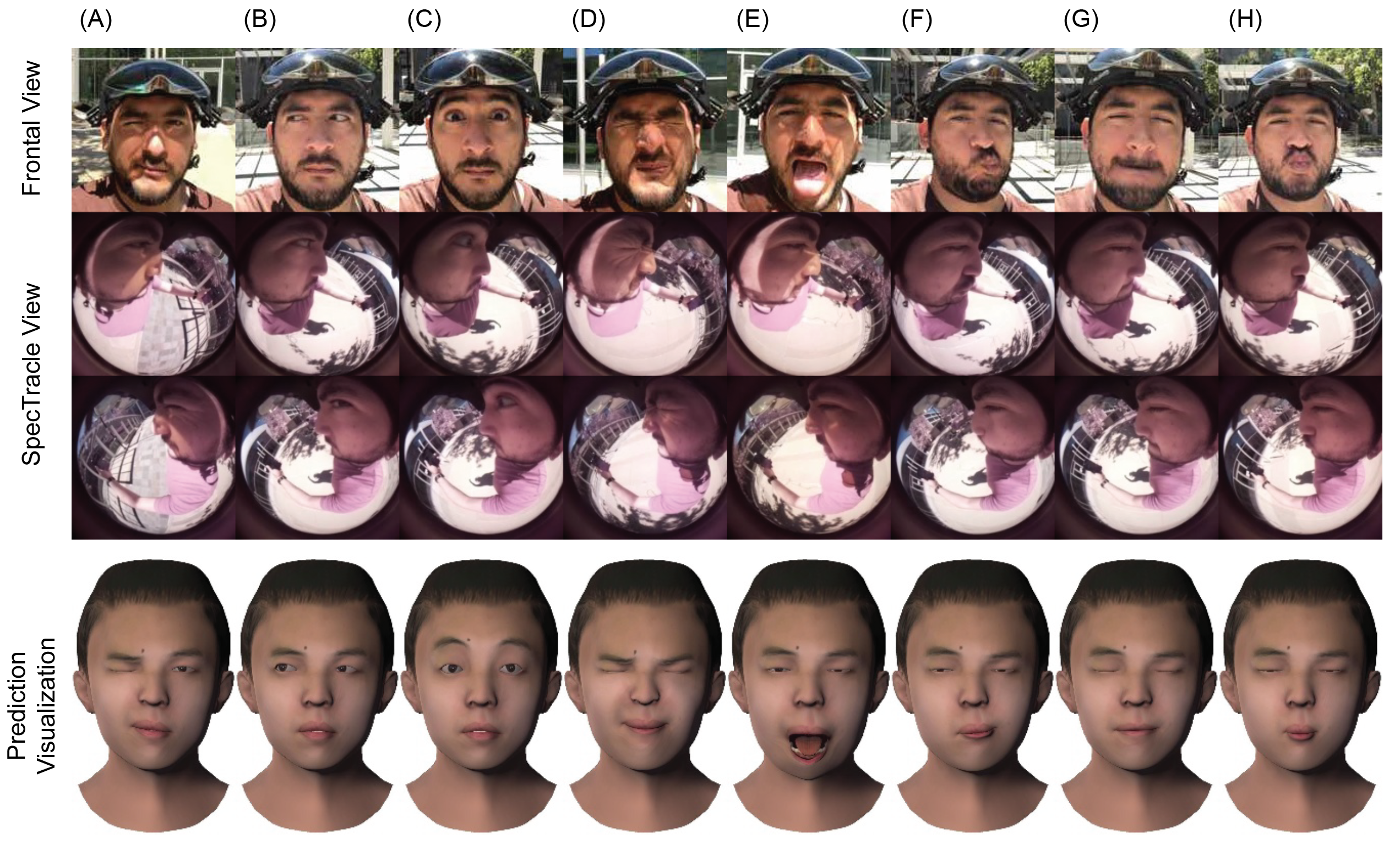}
  \caption{Sample visualizations of different facial expressions tracked using the two side cameras. (A) Blink single eye  with mouth tilted, (B) Look to the side, (C) Raise eyebrow, (D) Squeeze eye, (E) Stick out tongue with mouth open, (F) Kiss to the side, (G) Roll lips, (H) Pucker.}
  \label{visualized_examples}
\end{figure}

To visualize how the blend shape tracking maps to real facial expressions, we utilized our hand-rigged facial model to provide a more interpretable visualization of our results. In Figure \ref{visualized_examples}, we see that the system tracks different mixtures of expressions including movements of the eye, eyebrow, cheek, mouth, lip, and tongue.


\subsection{On-Device Processing}
We recognize the potential privacy concerns of using cameras in unconstrained environments. As can be seen in our sample image from the raw camera feed, the background can be seen. As the system would focus largely on face-to-face communication, the cameras would only be engaged during the use of facial motion tracking. In addition, we explored the processing runtime of the facial motion neural network on mobile GPUs on smartphones to estimate the runtime of the system in a real-time on-device processing pipeline. By running on local devices, the camera images were only captured temporarily, immediately processed for facial motion encodings, and discarded. 

We benchmarked our model's processing speed on a forward pass of a right/left image pair on two different mobile platforms:(1) Mobile GPU iPhone 11 and (2) Mobile GPU iPhone 7. To test how the model would perform in an online implementation, each batch in the forward pass includes one frame, which contains both the left and the right image. On the iPhone 7 with iOS version 13.7, for each feed forward, the model runs in 72 ms on average over 400 image pairs. On iPhone 11 with iOS version 13.7, each forward pass runs in 41 ms on average over 400 image pairs. (Table \ref{tab:benchmarks}) Our current camera capturing rate is around 8 fps, which is within the limit of the on-device neural network model inference ability.

\begin{table}[b]
  
  \begin{tabular}{ c|c|c|c}
    Platform & Mean Processing Time & Expected Frame Rate \\
    iPhone 7 Mobile GPU & 72ms & 13.9fps \\
    iPhone 11 Mobile GPU & 41ms & 24.4fps \\
\end{tabular}
\caption{Performance of our model on platforms it would need to run on.}
\label{tab:benchmarks}
\end{table}

\section{Discussion}

\subsection{User-Independent vs User-Calibrated Models}
Our current evaluation emphasizes testing in highly dynamic environments, where the subject moves around freely, in natural lighting conditions. It should be noted that the reported performance is likely to be close to an upper bound in terms of error, given that in many scenarios, the user would likely not move nearly as much as in our evaluation. During a small user study (N=3), we found that in a more static setting where the user is seated and not moving, we could achieve an average error as low as 0.76 (mm), compared to an error of 1.37 (mm) in our full evaluation of high movement, comparable to other works tested in controlled environments \cite{Chen2020cface}.  

When we compared the performance of the blend shape predictions between the user-independent model and the user-calibrated model in a location not introduced during calibration, the error of the prediction decreased from an average of 1.95 (mm) to 1.37 (mm). When tested in the location introduced during training, the error was further reduced to 1.17 (mm). In the accompanying video visualization for our work, we can see that compared to the user-independent model, the user-calibrated model predictions are more stable.

\subsection{Limitations}
\subsubsection{Differentiating Tongue and Inner Lip}
In some of our visualizations, we noted that the tongue out movement and lower lip shrug got confused both as tongue out. This makes sense given that the inner lip looks a lot like the tongue both in color and shape from the peripheral angle. 

Although we did not observe any specific issues, there is a possibility that ambient background color could sometimes lead to side-view images where the user's face appears to blend into the background, similar to how the tongue and lip get confused. Similarly, in poorly illuminated settings, such as the usage of the system at nighttime or in a dim room, the system may struggle to capture the face. These concerns motivate that an IR-based system may be desirable to mitigate concerns around major lighting or background issues. We note, however, that in the case of similar backgrounds, a few subjects wore tank top shirts, which exposed their shoulder such that the chin and the shoulder appeared similar in color, but the system did not perform worse in these situations.

\subsubsection{Under-represented User Groups}
Due to the COVID-19 pandemic, we were only able to recruit limited participants that consisted of lighter skin tone subjects (6 out of 9) and male subjects (6 out of 9). Although our current results did not show a significant difference in the model's performance across subjects, the results could be potentially improved by a more diverse dataset.

Furthermore, an observation we made during our system development is that the side facial images look similar between subjects of the same ethnicity. This was also true for our subjects who were of the same ethnicity but different genders. Based on the authors' observations, it is actually difficult to tell whether the images come from the same person or different people. Although we did not test this, it is possible that such similarity may have helped our generalized cross-validation performance. This observation potentially suggests that the user-independent model could perform even more favorably without user calibration with more training from people of similar ethnicity, but also motivates that the system may perform less reliably if the user-independent model does not incorporate users from different facial features. 

\subsubsection{Jitter in Tracking when Walking}
We observed that when the user walked around, although the system still generally tracked major movements, the stability of the predicted values jittered across frames. This results in what looks like a slight movement in the jaw, quivering lips, and a slight twitch in the eye. This issue is not present when standing or sitting still. This may be caused by changes in the shadows cast onto the face.

\subsection{Privacy}
Since the images from the glasses capture both the user's face and also the surroundings, the raw data reveals private information inadvertently. Therefore, we envision that rather than the computation being executed on a cloud-based service between users, the tracking system may need to run on the client side. As the AR system would likely not be powered by a full-fledged GPU, we benchmarked our system on a mobile GPU on a phone. One may imagine that such a GPU would be in an AR system or glasses may be interfaced with web services through a smartphone. 

In our benchmark, our model can run natively on an iPhone 11 at under 41ms per frame, a result that is yet optimized. Even though our neural network architecture is fairly complex, it runs fast enough to be used in a real-time system. 



\subsection{Future Directions}
During our data collection, we observed that our cameras were able to capture the movement of limbs, and therefore also contained information about the user's body posture. Thus, it's possible that our system can potentially go beyond facial tracking and predict body postures, which is another important part of face-to-face communication. For example, a wearer moves his or her hand onto or near his or her face. Expressions such as rubbing one's cheek or chin ultimately allow for a more immersive experience when these expressions can be translated into a virtual reality environment.

Beyond body tracking, the camera positioning lends itself to capturing activities around the face and around the body. For example, as similarly demonstrated by \citet{Bedri2020FitByte}, such a camera angle could be used to capture food and fluid intake. The camera positioning is also similar to that of pupil tracking systems. As mentioned earlier in the end-user system section, a potential way to integrate the system would be to double it with pupil tracking cameras in smart glasses. With a small adjustment of the lens to a wide-angle fisheye lens, we can expand pupil tracking cameras to perform full face tracking. 

\section{Conclusions}

This paper demonstrates that we are able to use two cameras mounted on the side of AR glasses to efficiently track a wide variety of different facial motions without sacrificing wearability. We also show that a neural network model trained on a diverse dataset is able to predict a new user's data. In addition, a quick user-specific calibration can be performed under 2 minutes to significantly increase the tracking performance. With the potential of modern mobile devices' computing power, our finding provides a new possibility to track facial motions in a mobile and unobtrusive manner for smart glasses and AR systems. 

\vspace{6pt} 

\supplementary{The following supporting information can be downloaded at: \href{https://www.youtube.com/watch?v=l_CdU9326-c}{Video S1: Demonstration of System Performance and Collected Data} .}


\authorcontributions{Conceptualization, E.W.; methodology, Y.X., V.V. and O.B.; software, Y.X., V.V., O.B. and J.E.; validation, Y.X., V.V., S.C. and J.E.; formal analysis, Y.X.; investigation, Y.X. and S.C.; resources, E.W.; data curation, Y.X. and S.C.; writing---original draft preparation, Y.X., V.V. and E.W.; writing---review and editing, Y.X., V.V., S.C., O.B., J.E. and E.W.; visualization, Y.X.; supervision, Y.X. and E.W.; project administration, E.W.; funding acquisition, E.W. All authors have read and agreed to the published version of the manuscript.}

\funding{This research received no external funding}

\institutionalreview{The study was conducted in accordance with the Declaration of Helsinki, and approved by the Institutional Review Board of the University of California, San Diego (Project \#200201XL, 05/12/2021).}

\informedconsent{Informed consent was obtained from all subjects involved in the study.}

\dataavailability{The data that support the findings of this study are available from the corresponding author upon reasonable request. } 

\acknowledgments{We thank all the participants who participated in the data collection. We also thank a team of students (Taylor Gulrajani, Nickan Shabdar, Huening Tong, Alistair Vizuet) from UCSD Mechanical and Aerospace Engineering Department for helping design the camera mount.}

\conflictsofinterest{The authors declare no conflict of interest.} 



\abbreviations{Abbreviations}{
The following abbreviations are used in this manuscript:\\

\noindent 
\begin{tabular}{@{}ll}
HMD & Head-Mounted Display\\
AR & Augmented reality\\
fps & Frames per second\\
VR & Virtual reality\\
MoCap & Motion capture\\
MR & Mixed reality
\end{tabular}
}




\begin{adjustwidth}{-\extralength}{0cm}

\reftitle{References}


\bibliography{sample-base}

\PublishersNote{}
\end{adjustwidth}
\end{document}